\documentclass[apj]{emulateapj}
\usepackage{hyperref}
\usepackage{epsfig}
\usepackage[normalem]{ulem}
\usepackage[utf8]{inputenc}
\usepackage[T1]{fontenc}
\usepackage{graphicx}
\usepackage{captcont}
\usepackage[english]{babel}
\usepackage{amssymb}
\usepackage{float}
\usepackage{multirow}
\usepackage{color}
\usepackage{threeparttable}
\usepackage{color}
\usepackage{arydshln}

\def\mearth{{\rm\,M_\oplus}}

\usepackage{times}
\slugcomment{}

\shorttitle{'Oumuamua as an extinct planetesimal fragment}
\shortauthors{Raymond, Armitage, \& Veras}

\begin{document}

\title{Interstellar object 'Oumuamua as an extinct fragment of an ejected cometary planetesimal}


\author{Sean N. Raymond}
\affil{ Laboratoire d'astrophysique de Bordeaux, Univ. Bordeaux, CNRS, B18N, allée Geoffroy Saint-Hilaire, 33615 Pessac, France; rayray.sean@gmail.com}

\author{Philip J. Armitage}
\affil{JILA, University of Colorado and NIST, 440 UCB, Boulder, CO 80309-0440, USA \\
Department of Astrophysical \& Planetary Sciences, University of Colorado, Boulder, CO 80309-0391, USA}

\author{Dimitri Veras}
\affil{Department of Physics, University of Warwick, Coventry CV4 7AL, UK \\
Centre for Exoplanets and Habitability, University of Warwick, Coventry CV4 7AL, UK}

\begin{abstract}
'Oumuamua was discovered passing through our Solar System on a hyperbolic orbit. It presents an apparent contradiction, with colors similar to those of volatile-rich Solar System bodies but with no visible outgassing or activity during its close approach to the Sun. Here we show that this contradiction can be explained by the dynamics of planetesimal ejection by giant planets. We propose that 'Oumuamua is an extinct fragment of a comet-like planetesimal born in a planet-forming disk that also formed Neptune- to Jupiter-mass giant planets. On its pathway to ejection 'Oumuamua's parent body underwent a close encounter with a giant planet and was tidally disrupted into small pieces, similar to comet Shoemaker-Levy 9's disruption after passing close to Jupiter. We use dynamical simulations to show that 0.1-1\% of cometary planetesimals undergo disruptive encounters prior to ejection. Rocky asteroidal planetesimals are unlikely to disrupt due to their higher densities. After disruption, the bulk of fragments undergo enough close passages to their host stars to lose their surface volatiles and become extinct. Planetesimal fragments such as 'Oumuamua contain little of the mass in the population of interstellar objects but dominate by number. Our model makes predictions that will be tested in the coming decade by LSST.
\end{abstract}

\section{The first interstellar object}
The discovery of interstellar objects was anticipated for decades~\citep{mcglynn89,kresak92,moromartin09,cook16,engelhardt17} yet 'Oumuamua was still a surprise. Its large brightness fluctuations point to an elongated shape, with a longest-to-shortest axis ratio of 5-10~\citep{meech17,bannister17,fitzsimmons18} (although a very close binary configuration has not been ruled out~\citep{gaidos18}). 'Oumuamua's lightcurve does not perfectly repeat, which indicates that the object is not spinning around any of its main axes but rather is tumbling~\citep{fraser18,drahus17}.  'Oumuamua's spectrum and colors fall within a broad range defined by primitive, volatile-rich Solar System bodies~\citep{meech17,jewitt17,masiero17,ye17,bannister17,fitzsimmons18}. The lack of outgassing during its closest approach to the Sun (perihelion distance of 0.25 AU) suggests that the upper $\sim 50$~cm of 'Oumuamua is depleted in volatiles, although its interior could be volatile-rich~\citep{fitzsimmons18}. 

The final stages of planet formation are relatively inefficient.  Planetesimals, the building blocks of planets, form from concentrations of small particles in gas-dominated planet-forming disks around young stars~\citep{chiang10,johansen14}. Giant planets first grow cores of several Earth masses ($\mearth$), then gravitationally accrete gas from the disk~\citep{pollack96}. Gas giants undergo a phase of runaway gas accretion whereas ice giants never reach the critical threshold for runaway~\citep{helled14}. 

A simple explanation for 'Oumuamua's origin is that it is a planetesimal that was ejected from another planetary system~\citep{laughlin17,raymond18}. Planetesimals are ejected by giant planets during gas accretion~\citep{raymond17}, subsequent dynamical clearing~\citep{duncan87,charnoz03}, and especially during instabilities among the giant planets. The broad eccentricity distribution of giant exoplanets~\citep{udry07b,butler06} is naturally reproduced if the observed planets are the survivors of system-wide instabilities during which giant planets repeatedly scattered off of each other, eventually leading to the ejection of one or more planets~\citep{adams03,chatterjee08,juric08,raymond10}.  Instabilities put giant planets in dynamical contact with a broad swath of remnant planetesimals, a majority of which are ejected~\citep{raymond10,raymond11,marzari14,raymond18}.

Here we demonstrate that 'Oumuamua is likely a fragment of a planetesimal that was torn apart and then dehydrated before its ejection. We analyze a large suite of dynamical simulations (Table 1) to show that 0.1-1\% of planetesimals pass so close to a giant planet that they should be torn to shreds. The majority of fragments undergo subsequent close passages to the host star before being ejected and should lose their surface volatiles and become extinct~\citep{nesvorny17}.

\section{Dynamical simulations}
Our analysis is based on several sets of already-published simulations~\citep{raymond09b,raymond10,raymond11,raymond12,raymond13b} whose properties are listed in Table 1. We now briefly summarize the setup. 

Each of our more than 5000 simulations started with three giant planets (gas giants, ice giants or a combination of both) and an outer disk of planetesimals orbiting a Solar-mass star. The innermost giant planet was placed on a near-circular, very low (but not zero) inclination at 5 AU. Subsequent planets were placed on exterior orbits, spaced by 4-5 mutual Hill radii so as to be in a marginally stable configuration~\citep{chambers96,marzari02,chatterjee08} close to strong mean motion resonances~\citep{raymond10}. The outermost planet's orbital radius was typically between 9 and 12 AU for the mass distribution used in the {\tt mixed} set of simulations and between 7 and 10 AU for the distribution used in the {\tt mixed2} simulations (Table 1). Each planet's bulk density was set to be ${\rm 1.3 \ g \ cm^{-3}}$, appropriate for Uranus, Neptune and Jupiter but not for Saturn (see Fig.~\ref{fig:disrupt}).  A 10 AU-wide disk of planetesimals was placed exterior to the giant planets, with its inner edge 2 Hill radii exterior to the outermost giant planet. Planetesimals were laid down following an $r^{-1}$ surface density profile, appropriate for the outer parts of observed protoplanetary disks~\citep{williams11} and similar in mass and radial extent to the primordial Kuiper belt~\citep{nesvorny12,deienno17}. Planetesimals were given small initial eccentricities and inclinations to avoid a 2-dimensional setup. A subset of simulations included additional material interior to the giant planets in the form of 500 additional planetesimals and 50 planetary embryos.~\citep{raymond11,raymond12}  Again this material was laid down following an $r^{-1}$ surface density profile, and these simulations were far more costly as a much shorter integration timestep was required.  However, we did not focus on `asteroidal' planetesimals in this paper because, as we demonstrate below, they are unlikely to represent the source of most interstellar objects.

Each simulation was integrated for 100-200 million years using the hybrid integrator in the Mercury code~\citep{chambers99}. Planets (and, in relevant simulations, planetary embryos) interacted gravitationally with all other particles in the simulation, but planetesimal particles did not self-gravitate. Collisions were treated as inelastic mergers and particles were considered ejected once they passed beyond 100 AU (except for the {\tt mini-OC} set, which had an ejection radius of $10^5$ AU). We discarded a small fraction of simulations that did not adequately conserve energy; this is why many sets in Table 1 do not have a round number of simulations.

Our simulations ejected a total of more than 3 million planetesimals.  Planetesimals were perturbed onto planet-crossing orbits either by having their eccentricities excited (e.g., by resonances) or because a giant planet was kicked onto an orbit that crossed a portion of the planetesimal disk.  Planetesimals underwent passages within the Hill radius of one or more giant planets before being imparted enough orbital energy to be ejected. The number of encounters is a strong function of the planet mass: in systems with only ice giants planetesimals underwent 1-2 orders of magnitude more close encounters before ejection than in systems only containing Jupiter-mass planets (e.g., the median number of encounters in systems with three $30 \mearth$ planets was 427 but in systems with three $3 M_{Jup}$ planets it was only 10).  The duration of instabilities is likewise a strong function of the planets' masses~\citep{raymond10}.  

\section{Tidal disruption and extinction of planetesimals}
On the pathway to ejection, some planetesimals pass so close to a giant planet that they may be torn apart by tidal forces. The critical radius for tidal disruption is a function of a planetesimal's density and tensile strength~\citep{cordes08,veras14,bear15,veras16}. We adopt a simple formalism for the tidal disruption radius $R_{tidal}$:
\begin{equation}
R_{tidal} \approx r \left(\frac{M_{giant}}{m}\right)^{1/3}  \approx {\rm 76,800 \  km} \left(\frac{\rho}{\rm 1 \ g \ cm^{-3}}\right)^{-1} \left(\frac{M_{giant}}{M_{Jup}}\right)^{1/3},
\end{equation}
\noindent where $m$ and $r$ are, respectively, a planetesimal's mass and radius, $M_{giant}$ is the giant planet's mass, $\rho$ is the planetesimal density and $M_{Jup}$ is Jupiter's mass. Given Jupiter's radius of $\sim$70,000~km, $R_{tidal}$ is only $\sim$10~\% above the planet's surface for $\rho = {\rm 1 \ g \ cm^{-3}}$.  A planetesimal must pass within a narrow range of encounter radii to be disrupted without colliding with Jupiter (see discussion below).

\begin{figure}
  \begin{center} 
  \leavevmode \epsfxsize=0.5\textwidth\epsfbox{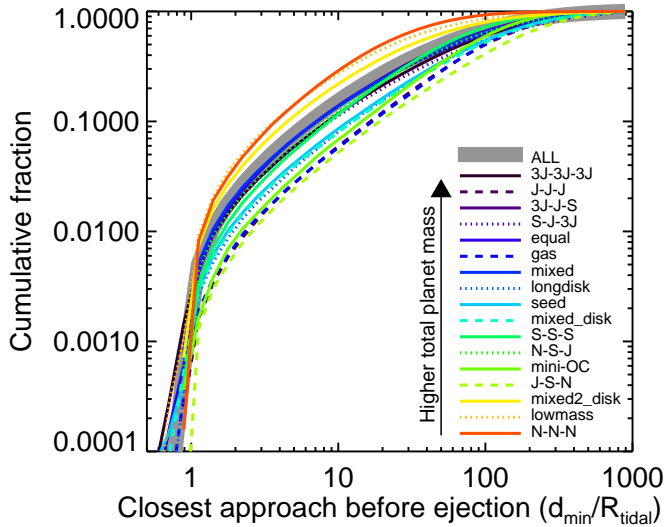}
    \caption[]{Cumulative distribution of the closest approach suffered by planetesimals on the pathway to ejection in different sets of simulations. The closest approach distance $d_{min}$ is normalized to the tidal disruption radius $R_{tidal}$, calculated assuming planetesimals of bulk density ${\rm 1 g \ cm^{-3}}$. The colors of different sets of simulations are roughly linked with their total planet masses, from most massive (black/purple/blue) to least massive (green/yellow/red). Only `cometary' planetesimals that originated past the giant planets' orbits are included. There is about an order of magnitude range between simulations in the fraction of potentially-disrupted planetesimals (see Table 1). } 
     \label{fig:minrt}
    \end{center}
\end{figure}

Figure~\ref{fig:minrt} shows the distribution of planetesimals' closest approach to a giant planet prior to ejection $d_{min}$, normalized to $R_{tidal}$ calculated for planetesimals with bulk density ${\rm 1 g \ cm^{-3}}$. Averaged over all simulations, 0.06\% / 0.8\% / 1.5\% of planetesimals passed within 1 / 1.5 / 2 $R_{tidal}$ before being ejected. However, there were variations between sets of simulations (see Table 1). In systems with lower-mass giant planets (yellow/red in Fig.~\ref{fig:minrt}), planetesimals tend to have closer closest approaches than in systems with high-mass giant planets (darker colors). Lower-mass planets require a larger number of encounters for ejection. Because the geometry of individual encounters is random, a higher number of encounters implies more ``rolls of the dice'' in lower-mass systems. While a larger number of encounters does not affect the typical encounter distance, on average it does lead to closer {\it closest} encounters before ejection for lower-mass systems. Among all sets of simulations, between 0.001-0.15\% / 0.26-1.9\% / 0.5-3.6\% planetesimals passed within 1 / 1.5 / 2 $R_{tidal}$ before being ejected (Table 1). The collisional timescale for the tidal debris should in most cases be much longer than the timescale for ejection~\citep[calculated using eqn 16 from][]{wyatt08}, so little collisional evolution of the debris stream is expected.

After repeated passages close to the Sun comets lose their surface volatiles. There exists a population of extinct comets on orbits similar to long-period and Halley-type comets called the Damocloids~\citep{jewitt05}. Dynamical models to match the relative populations of different Solar System comets employ a key parameter: the number of close passages to the Sun before a comet becomes extinct~\citep{levison97,disisto09,rickman17}.  \cite{nesvorny17} defined this as the number of orbits with perihelion $q < 2.5$~AU. To fit the populations of Halley-type and ecliptic comets, they found a critical number of close passages of $\sim 500-1000$.  However, to match the ratio of new-to-returning Oort cloud comets the critical number of passages within 2.5 AU could not be larger than 10. \cite{nesvorny17} attributed this to new Oort cloud comets generally being smaller than ecliptic or Halley-type comets, also proposed by \cite{brasser13b}.  

Planetesimals that go extinct before being ejected from their host stars should have surfaces reminiscent of volatile-rich bodies but without activity or outgassing (just like 'Oumuamua~\citep{meech17,jewitt17,fitzsimmons18}). Following \cite{nesvorny17} we assume that planetesimals cease being active after 1000 orbits with $q < 2.5$~AU. For planetesimal fragments we adopt \cite{nesvorny17}'s value of 10 orbits with $q<2.5$~AU to become inactive. We also test the effect of these assumptions.

\begin{figure}
  \begin{center} \leavevmode \epsfxsize=0.5\textwidth\epsfbox{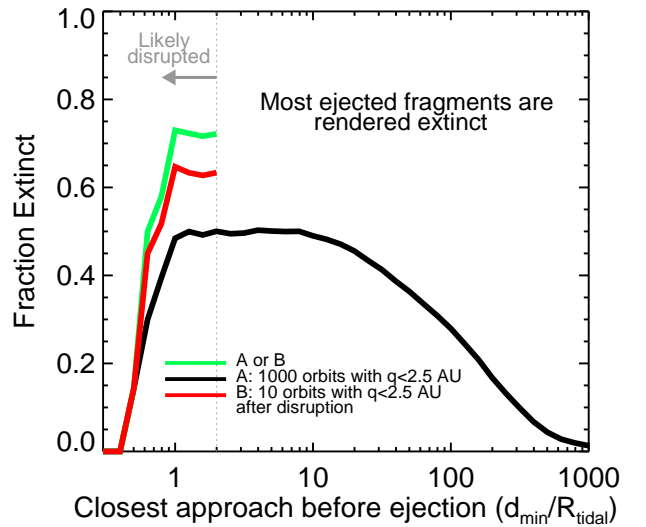}
    \caption[]{Fraction of planetesimals that undergo sufficient close pericenter passages to their host star to become extinct before being ejected, using the criteria from \cite{nesvorny17}. Given the low densities of observed comets, we consider encounters out to $2 R_{tidal}$ to have the potential for disruption (here, $R_{tidal}$ was calculated assuming a planetesimal density of ${\rm 1 g\ cm^{-3}}$). } 
     \label{fig:extinct}
    \end{center}
\end{figure}

Almost a third (32\%) of all planetesimals ejected in our simulations underwent at least 1000 passages within 2.5 AU of their host stars and should have been extinct. However, the fraction of extinct planetesimals increases for closer closest approach distances (Fig.~\ref{fig:extinct}). The large gravitational kicks required for a cometary planetesimal to enter the inner Solar System correlate with the close encounter distance: a closer encounter imparts a stronger velocity perturbation that generates a larger orbital deviation and an increased chance of entering the inner Solar System. 

The bulk (72\%) of potentially-disrupted planetesimals are rendered extinct before being ejected (Fig.~\ref{fig:extinct}).  They either completed at least 10 orbits with $q < 2.5$~AU after undergoing an extremely close approach to a giant planet (63\% of objects) or underwent a total of 1000 orbits with $q < 2.5$~AU (49\%) before being ejected. Extinct fragments thus dominate over still-active fragments. 

The criteria for extinction are only modestly-well constrained for Solar System comets and may also vary for different stellar types. For criteria of 100 / 1000 / $10^4$ passages within 2.5~AU, the fraction of planetesimals rendered extinct before ejection was 9.5\% / 32 \% / 51 \%. Yet fragments are more weakly affected by the extinction criterion. For a threshold of 1-100 orbits, between 54\% and 65\% fragments were rendered extinct after disruption and a total of between 64\% and 74\% of ejected disrupted objects were rendered inactive.  

Many of 'Oumuamua's peculiarities may naturally be explained if it is an extinct piece of a once volatile-rich planetesimal.  As a cometary fragment, 'Oumuamua should have spectral colors similar to volatile-rich Solar System bodies~\citep{meech17,jewitt17,fitzsimmons18}. However, as most fragments are rendered extinct before ejection, 'Oumuamua should not show any activity. The origin of 'Oumuamua's stretched-out shape~\citep{meech17,jewitt17,bannister17,bolin18,fitzsimmons18} and tumbling rotation~\citep{drahus17,fraser18} remain debated~\citep{domokos17,katz18,hoang18}, and we naively wonder whether a tidal disruption event could play a role in explaining those attributes.

\begin{figure}
  \begin{center} \leavevmode \epsfxsize=0.5\textwidth\epsfbox{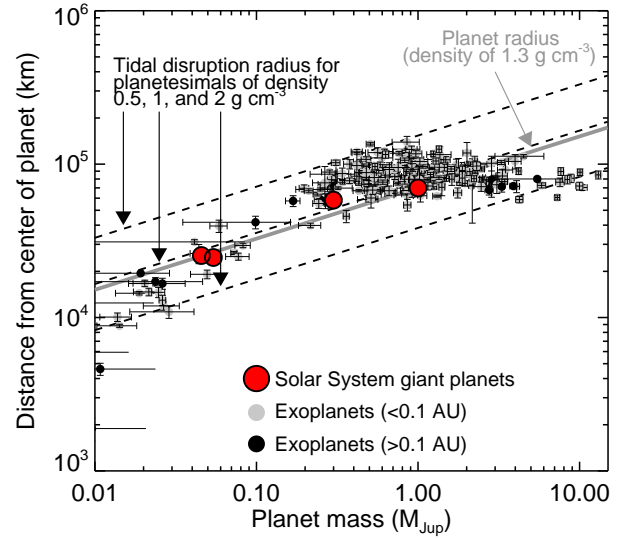}
    \caption[]{The parameter space available for tidal disruption of planetesimals, which requires that a planet's size is smaller than the tidal disruption radius $R_{tidal}$. Dashed lines represent $R_{tidal}$ for planetesimals with bulk densities of 0.5, 1 and 2 ${\rm g \ cm^{-3}}$ (top to bottom). The solid grey line is the planet radius assuming a constant density of ${\rm 1.3 g\ cm^{-3}}$ as in our simulations. The Solar System's giant planets and exoplanets with known masses and radii (including error bars) are also shown, divided into those on orbits closer or farther than 0.1 AU from their host stars (data downloaded from exoplanets.org~\citep{wright11} on Jan 30 2018). } 
     \label{fig:disrupt}
    \end{center}
\end{figure}

A number of factors make it difficult to estimate the true fraction of disrupted ejected planetesimals (Fig.~\ref{fig:disrupt}). For disruption to be possible, the perturbing planet must be smaller than the tidal disruption radius. Gaseous planets contract as they cool~\citep{fortney07,fortney10} but the tidal disruption radius only depends on the planet mass and so remains constant. The parameter space available for disruptive encounters is therefore a function of time. Our analysis may thus overestimate the rate of planetesimal disruption at earlier times, when many planetesimals are likely to be ejected~\citep{charnoz03,raymond17}.  On the other hand, we likely overestimate the densities of volatile-rich planetesimals. In Figs.~\ref{fig:minrt} and~\ref{fig:extinct} $R_{tidal}$ was calculated assuming a bulk density of ${\rm 1 g\ cm^{-3}}$. However, the densities of several cometary nuclei have been measured with spacecraft or inferred from theoretical studies are generally ${\rm \sim 0.5g\ cm^{-3}}$.~\citep{asphaug94,davidsson04,davidsson06,ahearn05,carry12,patzold16} Given that $R_{tidal}$ scales inversely with planetesimal density (Eq. 1), this implies a factor of two increase and justifies our inclusion of objects out to $2 R_{tidal}$ as potential disruptors (Fig.~\ref{fig:extinct}). In fact, comet Shoemaker-Levy 9 disrupted after passing 1.33 $R_{Jup}$ from Jupiter~\citep{sekanina94,weaver95,movshovitz12}, well beyond the ${\rm 1 g\ cm^{-3}}$ value for $R_{tidal}$ of $1.1 R_{Jup}$. The vast majority of giant exoplanets with measured masses and radii are capable of disrupting cometary planetesimals (Fig.~\ref{fig:disrupt}), even though many cannot efficiently eject planetesimals.~\citep{raymond10,laughlin17}

Rocky asteroidal planetesimals are unlikely to be progenitors of most interstellar objects for several reasons.  First, given that the snow line is typically located at a few AU~\citep{lecar06,martin12}, the vast majority of disks (by both surface area and mass) is cold enough to incorporate water ice and should thus be `cometary' rather than `asteroidal'.  Second, cometary bodies are more efficiently ejected by giant planets~\citep{raymond11,raymond12,raymond18} (asteroidal bodies are more efficiently driven into their stars~\citep{raymond11}).  Third, asteroids have higher densities than comets and correspondingly smaller tidal disruption radii. S- and C-complex asteroids have densities of ${\rm 2.5-3.5 g \ cm^{-3}}$ and ${\rm 1-2 g \ cm^{-3}}$, respectively~\citep{carry12}. Only very dense planets can disrupt such dense planetesimals (Fig.~\ref{fig:disrupt}). 

\section{Discussion}
Our simulations form the basis of a simple model for the population of interstellar objects. The occurrence rate of gas giants is 10-20\% for Sun-like stars~\citep{mayor11,winn15} but lower for low-mass stars~\citep{johnson07,lovis07}. Yet wide-orbit super-Earths and ice giants appear to be common around stars of all masses~\citep{beaulieu06,kubas12}.  Gas giant systems can eject tens of Earth masses of planetesimals and fragments (as in our Solar System~\citep{nesvorny12,deienno17}). Ice giant systems are unlikely to eject as much mass, perhaps $\sim 10 \mearth$ per system. Assuming a number-weighted gas giant frequency of a few percent and a 50\% occurrence rate for ice giants, $5-10 \mearth$ is ejected from each star, in line with estimates based on 'Oumuamua's detection~\citep{trilling17,laughlin17,raymond18,rafikov18,do18}, which are uncertain given the unknown underlying size distribution~\citep{raymond18}.  Ejection from binary stars~\citep{smullen16,sutherland16,jackson17} and from stars evolving off the main sequence~\citep{veras11,veras14b,stone15,hansen17,rafikov18} may also contribute to the population of interstellar objects. Given the uncertainties discussed above we roughly estimate that 0.1-1\% of planetesimals are disrupted prior to ejection. The size distribution of interstellar objects is comprised of a birth distribution of planetesimals~\citep{simon17,schafer17} overlaid with a distribution of fragments. As long as the typical fragment is smaller than the smallest planetesimals, fragments like 'Oumuamua dominate the ejected population although they contain little mass~\citep[see][for a derivation including the effect of the planetesimal size distribution]{raymond18}.

Our population model makes predictions that will be testable with a larger sample of interstellar objects. The population of interstellar objects should be dominated by fragments, in particular by small, extinct comet-like bodies similar to 'Oumuamua. Yet a modest fraction (roughly 1/4 to 1/3) of fragments should not have become inactive and should instead become active during their passage close to the Sun.  LSST will come online in the next few years and is estimated to find roughly 1 interstellar object per year~\citep{trilling17}, and the characteristics of these objects will directly constrain our model.  

\begin{deluxetable*}{c|cccc}
\tabletypesize{\scriptsize}
\tablewidth{0pt}
\tablenum{1}
\tablecaption{Simulation overview.  The columns represent (i) the name of the set used in the paper in which it was first presented; (ii) the number of simulations from that set used in this analysis; (iii) the mass distribution of the three giant planets in that set; (iv) the fraction of planetesimals that underwent an encounter closer than a given fraction of $R_{tidal}$ prior to ejection; and (v) the appropriate reference. Different sets of simulations varied either the giant planets' masses (as indicated) or other parameters: the {\tt gas} simulations included the effects of aerodynamic drag from the dissipating protoplanetary disk, the {\tt seed} simulations included a handful of Earth-mass bodies within the planetesimal disk, the {\tt longdisk} simulations included a 20 AU-wide (rather than 10 AU-wide) planetesimal disk, and the {\tt mini-OC} simulations had an ejection radius of $10^5$ AU (instead of 100 AU).}
\tablehead{ 
\colhead{Set}	&
\colhead{N (sims)}	&
\colhead{Giant Planet mass distribution}	&
\colhead{Fraction with $d_{min} < 1 / 1.5 / 2 \ R_{tidal}$} &
\colhead{Reference}
}
\startdata
mixed\_disk & 1000 & $dN/dM \propto M^{-1.1}$ from $1 M_{\rm Sat}$ to $3 M_{\rm Jup}$ & 0.08\% / 0.9\% / 1.8\% & 
~\cite{raymond09b,raymond10}\\
mixed2\_disk & 1000 & $dN/dM \propto M^{-1.1}$ from $10 \mearth$ to $3 M_{\rm Jup}$ & 0.07\% / 1.3\% / 2.7\% & 
~\cite{raymond09b,raymond10}\\
3J-3J-3J & 200 & three equal-mass planets of $3 M_{Jup}$ & 0.15\% / 0.8\% / 1.5\% & ~\cite{raymond10}\\
J-J-J & 498  & three equal-mass planets of $M_{Jup}$ & 0.025\% / 0.3\% 0.6\% & ~\cite{raymond10}\\
S-S-S & 499 & three equal-mass planets of $M_{Sat}$ & 0.04\% / 0.6\% / 1.2\% & ~\cite{raymond10}\\
N-N-N & 435 & three equal-mass planets of $30 \mearth$ & 0.016\% / 1.7\% / 3.2\% &~\cite{raymond10}\\
J-S-N & 250 & inner to outer: $M_{Jup}$, $M_{Sat}$, $30 \mearth$ & 0.001\% / 0.26\% / 0.5\% & ~\cite{raymond10}\\
N-S-J & 248 & inner to outer: $30 \mearth$, $M_{Sat}$ , $M_{Jup}$ & 0.06\% / 0.8\% / 1.4\% & ~\cite{raymond10}\\
3J-J-S & 218 & inner to outer: 3 $M_{Jup}$, $M_{Jup}$, $M_{Sat}$ & 0.08\% / 0.8\% / 1.7\% & ~\cite{raymond10}\\
S-J-3J & 250 & inner to outer: $M_{Sat}$, $M_{Jup}$, 3 $M_{Jup}$ & 0.08\% / 0.7\% / 1.5\% & ~\cite{raymond10}\\
mini-OC & 50 & three equal-mass planets of $M_{Jup}$ & 0.04\% / 0.3\% / 0.7\% & ~\cite{raymond13b}\\
mixed & 187 & $dN/dM \propto M^{-1.1}$ from $1 M_{\rm Sat}$ to $3 M_{\rm Jup}$ & 0.05\% / 0.5\% / 1.0\% & 
~\cite{raymond11,raymond12}\\
lowmass & 97 & $dN/dM \propto M^{-1.1}$ from $10 \mearth$ to $3 M_{\rm Jup}$ & 0.1\% / 1.8\% / 3.5\% & 
~\cite{raymond12}\\
equal & 240 & three equal-mass planets of $M_{Jup}$ & 0.06\% / 0.85\% / 1.6\% & ~\cite{raymond12}\\
seed & 91 & $dN/dM \propto M^{-1.1}$ from $1 M_{\rm Sat}$ to $3 M_{\rm Jup}$ & 0.04\% / 0.5\% / 1.0\% & 
~\cite{raymond12}\\
longdisk & 96 & $dN/dM \propto M^{-1.1}$ from $1 M_{\rm Sat}$ to $3 M_{\rm Jup}$ & 0.03\% / 0.5\% / 0.9\% & 
~\cite{raymond12}\\
gas & 38 & $dN/dM \propto M^{-1.1}$ from $1 M_{\rm Sat}$ to $3 M_{\rm Jup}$ & 0.03\% / 0.3\% / 0.6\% & 
~\cite{raymond12}\\
\hline 
TOTAL & 5397
\enddata
\end{deluxetable*}

\section*{Acknowledgments}
We are grateful to Noel Gorelick for allowing us to run the bulk of the simulations on computers at Google. D.V. acknowledges the support of the STFC via an Ernest Rutherford Fellowship (grant ST/P003850/1). P.J.A. acknowledges support from NASA through grant NNX16AB42G. S.~N.~R. thanks the Agence Nationale pour la Recherche for support via grant ANR-13-BS05-0003-002 (grant MOJO) and acknowledges NASA Astrobiology Institute's Virtual Planetary Laboratory Lead Team, funded via the NASA Astrobiology Institute under solicitation NNH12ZDA002C and cooperative agreement no. NNA13AA93A.


\end{document}